\documentclass{ws-procs975x65}

\begin{document}


%
%

\title{Electrically charged compact stars.}
\author{Subharthi Ray\footnote{Present Address: Inter
University Centre for Astronomy and Astrophysics, Post Bag 4, Pune
411007, India.} and Manuel Malheiro}
\address{Instituto de Fisica, Universidade Federal Fluminense,
Niteroi 24210-340 RJ, Brasil}
\author{Jos\'e P. S. Lemos}
\address{Centro Multidisciplinar de Astrof\'{\i}sica - CENTRA, Departamento
de F\'{\i}sica, Instituto Superior T\'ecnico, Av. Rovisco Pais 1,
1096 Lisboa, Portugal}
\author{Vilson T. Zanchin}
\address{Universidade Federal Santa Maria, Departamento de Fisica,
BR-97119900, Santa Maria, RS, Brazil}

\maketitle


\abstracts{We review here the classical argument used to justify
the electrical neutrality of stars and show that if the pressure
and density of the matter and gravitational field inside the star
are large, then a charge and a strong electric field can be
present. For a neutron star with high pressure ($\sim 10^{33}$ to
$10^{35}$ dynes /cm$^2$) and strong gravitational field ($\sim
10^{14}$ cm/s$^2$), these conditions are satisfied. The
hydrostatic equation which arises from general relativity, is
modified considerably to meet the requirements of the inclusion
of the charge. In order to see any appreciable effect on the
phenomenology of the neutron stars, the charge and the electrical
fields have to be huge ($\sim 10^{21}$ Volts/cm). These stars are
not however stable from the viewpoint that each charged particle
is unbound to the uncharged particles, and thus the system
collapses one step further to a charged black hole}


Majumdar\cite{maj47} and Papapetrou\cite{papa47} studied charged
dusts in the light of general relativity, and later also by many
others (see Ray et al.\cite{remlz03} for a detailed reference).
Charged fluid spheres have been studied by Bekenstein\cite{bek71},
Bonnor\cite{bon80}, Zhang et al.\cite{zhang82}, etc. Zhang et
al.\cite{zhang82} indirectly verified that the structure of a
star, for a degenerate relativistic fermi gas, is significantly
affected by the electric charge  just when the charge density is
close to the mass density.

We took trapped protons to be the charge carriers in the star. The
effect of charge does not depend on its sign by our formulation.
The energy density which appears from the electrostatic field
will {\it add up} to the total energy density of the system,
which in turn will  help in the {\it gaining} of the  total mass
of  the system. The modified Tolman-Oppenheimer-Volkoff (TOV)
equation now has extra terms due to the presence of  the
Maxwell-Einstein stress tensor. We solve the modified TOV
equation for  polytropic EOS assuming that the charge density
goes with the matter density.


The detailed relations are shown in Ray et al.\cite{remlz03}. We
used the modified TOV and see the effect of charge on a model
independent polytropic EOS. We assumed the  charge is proportional
to the mass density ($\epsilon$) like $\rho_{ch}=f \times
\epsilon $ where $\epsilon=\rho c^2$ is in [MeV/fm$^3$].

The polytropic EOS is given by $P=\kappa \rho^{1+1/n} $ where $n$
is the  polytropic index and for our present choice of EOS, we
took $n=1.5$ and $\kappa=0.05[fm]^{8/3}$.

\begin{figure}[ht]
\centerline{\epsfxsize=5cm\epsfbox{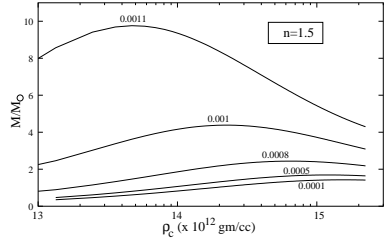}} \caption{Central
density against mass for different values of the factor
$f$.\label{fig:poly-e-m}}
\end{figure}


In  Fig.(\ref{fig:poly-e-m}), we plot the mass as function of the
central density, for  different values  of the  charge fraction
$f$. For the charge fraction $f=0.0001$, we do not see any
departure on the stellar structure from that of the chargeless
case. This value of $f$ is $critical$ because any increase in the
value beyond this, shows enormous effect on the structure. The
increase of the maximum mass of the star is very much non-linear,
as can be seen from the Fig. (\ref{fig:poly-e-m}).

\begin{figure}[htbp]
\centerline{\psfig{figure=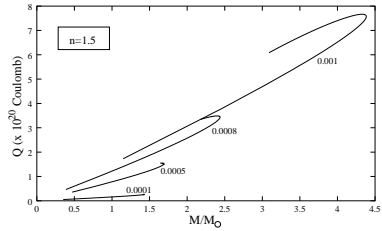,width=5cm}}
\caption{\label{fig:poly-q-m}The variation of the charge with
mass (right) for different $f$.}
\end{figure}

The Q$\times$ M diagram of Fig.(\ref{fig:poly-q-m}) shows the
mass of the stars against their surface charge. We have made the
charge density proportional to  the energy  density and so  it
was expected  that the charge, which is  a volume integral of the
charge  density, will go in the same  way as the  mass, which is
also a volume integral  over the mass density.  The slope of the
curves comes from the different charge fractions. If we consider
that the maximum allowed charge estimated by the condition ($U
\simeq \sqrt{8\pi P} < \sqrt{8\pi\epsilon}$) for $\frac{dP}{dr}$
to be negative (Eq.~12 of Ray et al.\cite{remlz03})),  we see that
the curve for the maximum charge  in Fig.(\ref{fig:poly-q-m}) has
a slope of 1:1 (in a charge scale of 10$^{20}$
Coulomb\cite{remlz03}). This scale can easily be understood if we
write the charge as
$Q=\sqrt{G}M_\odot\frac{M}{M_\odot}\simeq10^{20}\frac{M}{M_\odot}{\rm
Coulomb}.$ This charge Q  is the charge at  the surface of the
star where the  pressure and  also $\frac{dP}{dr}$  are zero.

The total  mass of the star increases with increasing charge
because the  electric energy density $adds~on$ to  the mass energy
density. This change in  the mass is low for smaller charge
fraction  and going  up to  12 times  the value  of chargeless
case  for maximum allowed charge  fraction $f=0.0011$. The most
effective term   in  Eq.~12 of Ray et al.\cite{remlz03} is  the
factor ($\rm M_{tot}+4\pi r^3P^*$). $P^*= P-\frac{U^2}{8\pi}$ is
the effective pressure of the system because the effect of charge
decreases the outward fluid pressure, negative in sign to the
inward gravitational pressure. With the increase of charge, P$^*$
decreases, and hence the gravitational negative part of Eq.~12 of
Ray et al.\cite{remlz03} decreases. So, with the softening of the
pressure  gradient, the system allows more radius for the star
until  it reaches  the surface where the pressure (and
$\frac{dP}{dr}$) goes to zero. Since $\frac{U^2}{8\pi}$ cannot
be  too much larger than the pressure in order to maintain
$\frac{dP}{dr}$  negative, so we have a limit  on  the charge,
which comes from the relativistic effects of the gravitational
force and not just only from the repulsive Coulombian part.

In our study, we have shown that a high density system like a
neutron star can hold huge charge of the order of 10$^{20}$
Coulomb considering the global balance of forces. With the
increase of charge, the maximum mass of the star recedes back to
a lower density regime. The stellar mass also increases rapidly
in the critical limit of the maximum charge content, the systems
can hold. The radius also increases accordingly, however keeping
the M/R ratio increasing with charge. The increase in mass is
primarily brought in by the softening of the pressure gradient due
to the presence of a Coulombian term coupled with the
Gravitational matter part. Another intrinsic increase in the mass
term comes through the addition of the electric energy density to
the mass density of the system. The stability of these charged
stars are however ruled out from the consideration of forces
acting on individual charged particles. They face enormous radial
repulsive force and leave the star in a very short time. This
creates an imbalance of forces and the gravitational force
overwhelms the repulsive Coulomb and fluid pressure forces and
the star collapses to a charged black hole. Finally, these
charged stars are supposed to be very short lived, and are the
intermediate state between a supernova collapse and charged black
holes\cite{remlz03}.

\end{document}